\documentclass[aps,prl,twocolumn,showpacs,superscriptaddress,twoside,showpacs]{revtex4-2}
\usepackage{graphicx}
\usepackage{dcolumn}
\usepackage{bm}
\usepackage{booktabs}
\usepackage{amsmath}
\usepackage{color}

\begin{document}

\title{Comment on ``Transverse Wobbling in $^{135}$Pr [Phys. Rev. Lett. 114, 082501 (2015)]"}

 \author{S. Guo}
\affiliation{CAS Key Laboratory of High Precision Nuclear Spectroscopy, Institute of Modern Physics, Chinese Academy of Sciences, Lanzhou 730000, China} 
\affiliation{School of Nuclear Science and Technology, University of Chinese Academy of Science, Beijing 100049, People's Republic of China}

 \date{}
\begin{abstract}

In [J. T. Matta et al., Phys. Rev. Lett. 114, 082501 (2015)] a transverse wobbling band was reported in $^{135}$Pr. The critical experimental proof for this assignment is the E2 dominated linking transitions between the wobbling and normal bands, which are supported by two experiments performed with Gammasphere and INGA. However, the M1 dominated character cannot be excluded based on the reported experimental information, indicating that the wobbling assignment is still questionable.

\end{abstract}



\maketitle

The present comment exposes some concerns on the reported experimental proofs on the declared wobbling band in $^{135}$Pr \cite{Matta}.



\begin{figure}[ht]
	\vskip -. cm
	\hskip -. cm
	\centering\includegraphics[clip=true,width=0.4\textwidth]{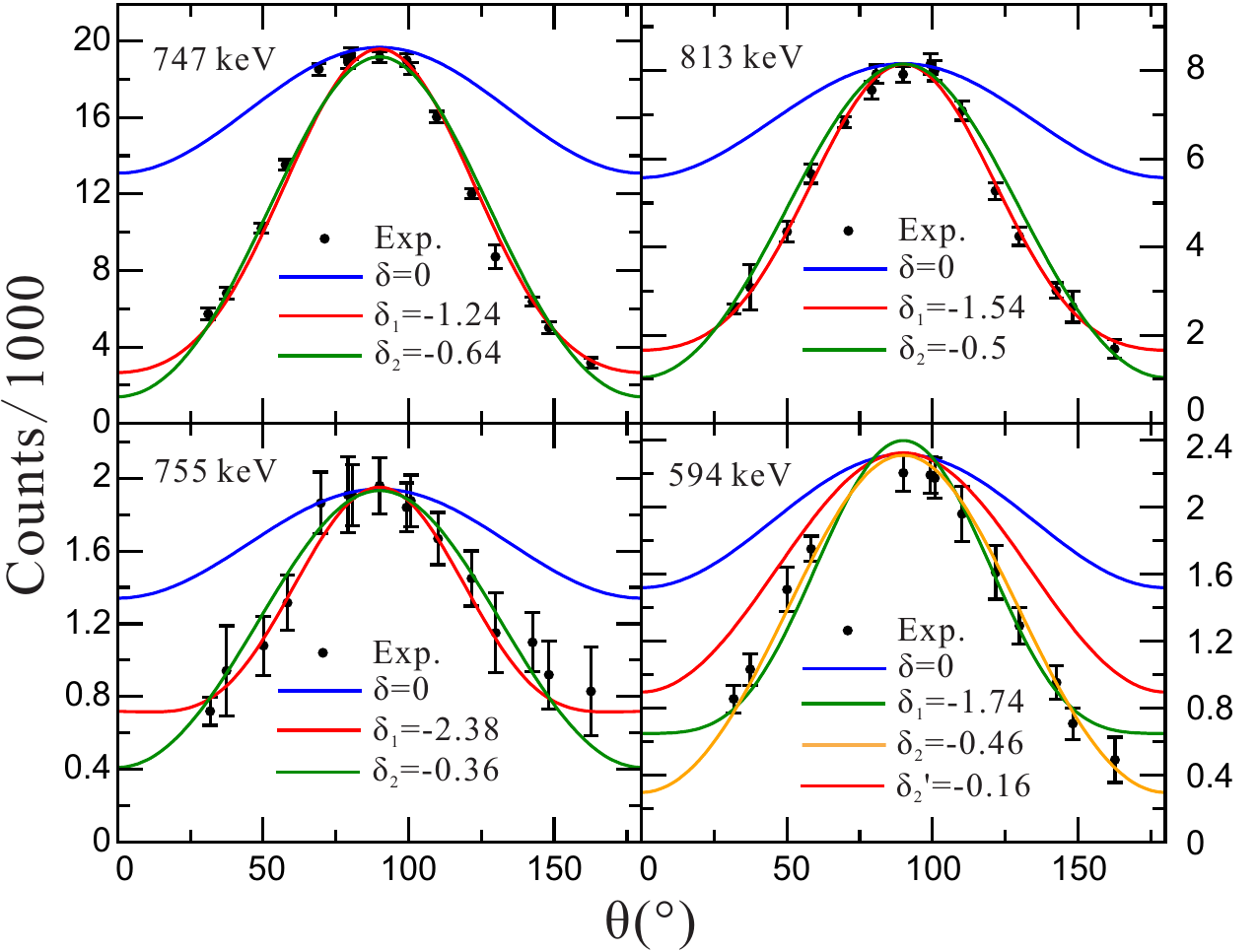}
	\vskip -0. cm
	\caption{(Color online) Estimated angular distribution curves and experimental results for the linking transitions in $^{135}$Pr. For the 594-keV transition in $^{135}$Pr, $\delta_2$ = -0.46 is deduced from the curve marked with $\delta_2 '$ = -0.16 in Ref. \cite{Matta}, therefore two corresponding curves are both plotted.}
	\label{fig1}
\end{figure}

Angular distribution curves can be estimated assuming appropriate values for the mixing ratio ($\delta$) and the alignment parameter ($\sigma/I$). From the comparison with the experimental results, it is possible to deduce suitable mixing ratios to get a good agreement between estimated and experimental results. Usually, equally good agreements can be achieved with two mixing ratios $|\delta_1|>1$ and $|\delta_2|<1$ (see Fig. \ref{fig1}). As the key experimental criterion to judge if a band is wobbling band, the mixing ratio should be deduced by comparing the two curves, with both $\delta_1$ and $\delta_2$, to the experimental results with convincing errors. However, in the commented article \cite{Matta}, for each $M1/E2$ transition, only one curve with either $\delta_1$ or $\delta_2$ was plotted and compared with pure $M1$ transitions ($\delta=0$), leading to a false belief that the correct solution for the mixing ratio is reliably established since the curve for $\delta=0$ is far apart from the experimental points. Based on the reported results from angular distribution measurements (Fig. 2 in Ref. \cite{Matta}), we added the curve with the other fitted solution to each figure (see the green curves in Fig. \ref{fig1}). Actually, both the two curves with $\delta_1$ and $\delta_2$ fit with the experimental points well within the reported errors.

\begin{figure}[ht]
	\vskip -. cm
	\hskip -. cm
	\centering\includegraphics[clip=true,width=0.4\textwidth]{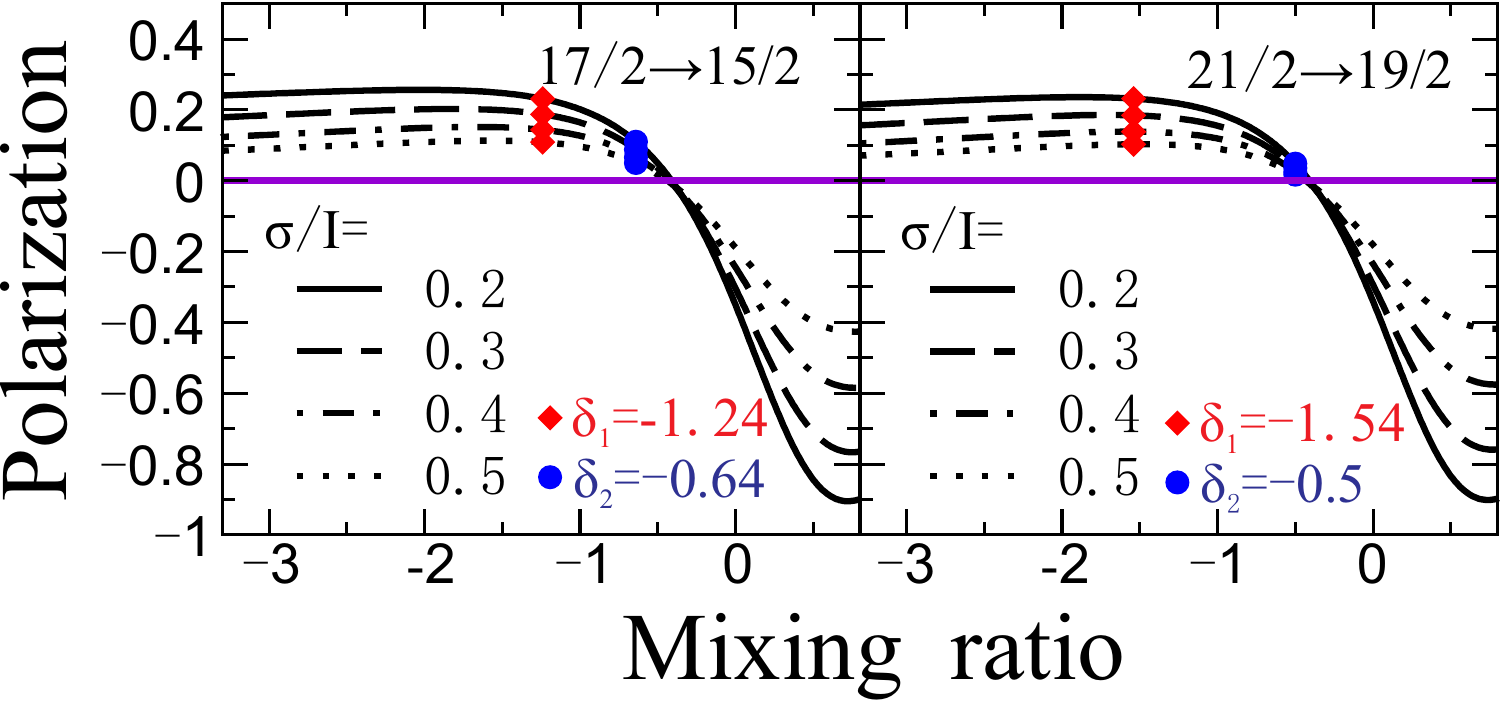}
	\vskip -0. cm
	\caption{(Color online) The polarization coefficients as functions of mixing ratios with different $\sigma$/I parameters.}
	\label{fig2}
\end{figure}

In the commented article \cite{Matta}, positive asymmetries were deduced from the linear polarization measurements, and were regarded as a clear proof to identify the two transitions of 747.0 and 812.8 keV as predominately electric in nature. However, a positive asymmetry does not exclusively imply a predominantly electric character for $\Delta$I=1 $M1/E2$ transitions. The polarization value, which should have the same sign with the asymmetry, can be estimated as a function of mixing ratio, with known spins of initial and final states, and assumed $\sigma$/I. For the two transitions of 747.0 and 812.8 keV, the estimated polarization values are positive with both $\delta_1$ and $\delta_2$ (see Fig. \ref{fig2}). Therefore both $\delta_1$ and $\delta_2$ cannot be ruled out by the reported experimental results in Ref. \cite{Matta}. Considering the polarization values are slightly different for $\delta_1$ and $\delta_2$, the real solution can only be decided by an extra precise analysis. To deduce the polarization values, the calibrated polarization sensitivity and deduced $\sigma/I$ parameters for each transition in the INGA measurement should be taken into account, which were not mentioned in the Ref. \cite{Matta}.

In summary, it is insufficient to declare a wobbling band based on the reported experimental proofs in Ref. \cite{Matta}. Further experimental researches are necessary to clarify the nature of this band.

The author thanks Professor C. M. Petrache for useful discussions. This work has been partly supported by the National Natural Science Foundation of China, under contract No. U1932137.

\end{document}